\documentclass[12pt]{article}
\usepackage{amssymb}
\usepackage{setspace}
\begin{document}

\newcommand{\bra}[1]{\langle #1|}
\newcommand{\ket}[1]{|#1\rangle}
\newcommand{\braket}[2]{\langle #1|#2\rangle}
\newcommand{\be}{\begin{eqnarray}}
\newcommand{\ee}{\end{eqnarray}}
\newcommand{\g}{\theta}
\newcommand{\co}[2]{\lbrack #1 , #2 \rbrack}
\newcommand{\Norm}[1]{\| #1 \|}
\newcommand{\norm}[1]{| #1 |}
\newcommand{\BigNorm}[1]{\Big\| #1 \Big\|}
\newcommand{\Bignorm}[1]{\Big| #1 \Big|}
\newcommand{\ex}[1]{\mathbf{E}\lbrack #1 \rbrack}
\newcommand{\p}{\partial}
\newcommand{\ncr}[2]{\Big( \begin{array}{c} #1 \\ #2 \end{array} \Big)}
\newcommand{\om}{\omega}
\newcommand{\mtr}[4]{\left(
 \begin{array}{cc} #1 & #2 \\ #3 & #4
 \end{array} \right)}
\newcommand{\id}{\mathbb{I}}
\newcommand{\mbold}[1]{\mbox{\boldmath ${#1}$}}

\def\maketitle{
   \begin{flushright}
   Nuc. Phys. \textbf{B722}, 249 (2005)\\ hep-th/0504009\\ DAMTP-2005-32
   \end{flushright}
   \begin{center}
   \Huge{Logarithmic Primary Fields in Conformal and Superconformal Field Theory}\\
   \vskip 0.7 cm
   \large{Jasbir Nagi}\\ \vskip 0.7 cm \large{J.S.Nagi@damtp.cam.ac.uk}\\ \vskip 0.7 cm \large{DAMTP,
   University of Cambridge, Wilberforce Road,}\\ \vskip 0.7 cm \large{Cambridge,
    UK, CB3 0WA}
   \vskip 0.7 cm
   \end{center}
   }

\maketitle
\begin{abstract}
In this note, some aspects of the generalization of a primary field to
the logarithmic scenario are discussed. This involves understanding
how to build Jordan blocks into the geometric definition of a primary field of a
conformal field theory. The
construction is extended to $N=1,2$ superconformal theories. For the
$N=0,2$ theories, the two-point functions are calculated. 
\end{abstract}
\newpage

\section{Introduction}
In recent years, Logarithmic Conformal Field Theories (LCFTs) have come under
much investigation\cite{flo}. The reasons include possible applications to statistical
physics \cite{piroux}, possible applications to WZW theories
\cite{gaber2}\cite{nich}, possible applications to D-Brane dynamics
\cite{mav1}\cite{mav2}, and a potential understanding of how to
control non-unitary quantum field theories. LCFTs are characterized by non-unitary
behaviour, such as logarithms in correlation functions and
indecomposable Jordan blocks in the Virasoro representation
theory.

Even though a LCFT has such recognizable characteristics, a clear cut
definition of a LCFT still does not really exist. Moreover, how the usual
machinery of CFT generalizes to the LCFT case is still not completely
understood. Many approaches to these problems have been followed
\cite{flo2}\cite{gaber3}. In this note, the construction of a LCFT in
terms of its primary fields will be analyzed, following
\cite{rahimi}\cite{khorrami}\cite{moghi}. This approach introduces
indecomposable Jordan blocks by hand, and essentially comes down
to modifying existing constructions by introducing nilpotent
`variables', which algebraically mimic the Jordan block
structure. 

In this note, many of the features of the bosonic are reviewed in a
slightly different light from the previous literature. In particular,
a more geometric approach is considered, and 
instead of nilpotent variables, Jordan blocks are used from the
outset. This approach is then generalized to the supersymmetric
$N=1,2$ cases.

Sections 2-4 describe the bosonic theory in a manner that naturally
generalizes to the supersymmetric case. In these sections, a primary
field is defined, and the infinitesimal transformations obtained. In
order to verify the differential operators obtained indeed give the
required primary field, the generators of
the infinitesimal transformation are exponentiated. As an
application, using global conformal symmetry, the two point function is calculated. These sections yield and extend some
results from \cite{rahimi}\cite{moghi}\cite{flo3}.

Using the machinery developed in sections 2-4, section 5 looks at the $N=1$
theory, obtaining and extending some results from \cite{khorrami},
although in a more geometric fashion. Since the bosonic part of the Cartan sub-algebra of
the $N=1$ theory is same as for the bosonic case, the machonery works
in much the same way.

Sections 6-8 look at the $N=2$ theory. Since the Cartan subalgebra is
larger than the $N=0,1$ theories, more Jordan blocks can potentially
appear. Section 7 defines a $N=2$ logarithmic primary field that
accounts for these extra Jordan Blocks. In order to study what
further logarithms might occur, the two point function is
calculated using global conformal symmetry.

\section{General Framework}
Consider the one-form $dz$, the matrix
\be
M = \left( \begin{array}{cc} h & 1 \\ 0 & h \end{array} \right) =
h\mathbb{I}_2 + J
\ee
and the formal one-form $dz^M$. Then, using $a^b = \exp (b \log a)$
and the series definition for $\exp$,
\be\label{n0oneform}
dz^M = dz^{h\mathbb{I}_2 + J} = dz^{h\mathbb{I}_2}dz^J =
\mathbb{I}_2dz^h\exp(J\log (dz)) = dz^h(\mathbb{I}_2 + J\log (dz))
\ee
A more proper definition of $dz^M$ might be $dz^h(\mathbb{I}_2 +
J\log (dz))$, although exactly what $\log(dz)$ means is not apparent
to the author. Let the matrix act on a column vector
\be
v = \left( \begin{array}{cc} \phi_0(z) \\ \phi_1(z) \end{array} \right)
\ee
and consider a conformal transformation $f:z\mapsto z'$. Then, under
pull-back, one has for $dz^Mv$
\be
&&f^*\Bigg( dz'^h \left( \begin{array}{cc} 1 & \log (dz') \\ 0 & 1
\end{array} \right)\left( \begin{array}{cc} \phi_0(z') \\ \phi_1(z')
\end{array} \right)\Bigg) \\  &&=\Big(\frac{dz'}{dz}\Big)^hdz^h\left(
\begin{array}{cc} 1 & \log \big(\frac{dz'}{dz}\big)+\log (dz) \\ 0 & 1
\end{array} \right) \left( \begin{array}{cc} (\phi_0\circ f)(z) \\
  (\phi_1\circ f)(z) \end{array} \right)\\ &&=\Big(\frac{dz'}{dz}\Big)^hdz^h\left(
\begin{array}{cc} \phi_0 + \log \big(\frac{dz'}{dz}\big)\phi_1 +\log (dz)\phi_1 \\ \phi_1
\end{array} \right)\\ && =: dz^h \left( \begin{array}{cc} \phi'_0(z) +
  \log(dz)\phi'_1(z)\\ \phi'_1(z) \end{array} \right)
\ee
This then gives the well known transformations for a logarithmic
primary field \cite{rahimi}.
It seems that $\log (dz)$, is well-defined in the
algebraic sense, up to the arbitrary phase
that can be added, although the notion of constructing a geometric
object out of $\log (dz)$ seems very unusual. Here, it has been assumed that $dz$
and $\log (dz)$ are linearly independent.

More generally, one could consider raising $dz$ to the power $h+J$,
where $J^n=0$, $J^{n-1}\neq 0$, $hJ=Jh$, and $h$ not nilpotent, which
has a unique (up to conjugation), faithful representation of smallest dimension
$=n$. Choosing the $n$-dimensional representation where $J$ has $1$s
just above the
leading diagonal, and is zero elsewhere, and going through the same
machinery, one finds
\be\label{gen1form}
dz^{h+J} = dz^h\sum_{i=0}^{n-1}\frac{J^i(\log dz)^i}{i!}
\ee
Now, $v$ is a column
\be
v_i = \phi_i(z),\qquad i=0,\ldots, n-1
\ee
and hence
\be
(dz^{h+J}v)_i = dz^h \sum_{j=0}^{n-i-1}\frac{1}{j!}\phi_{j+i}(z)(\log dz)^j
\ee
Pulling back gives
\be
f^*(dz'^{h+J}v)_i(z) &=&
\Big(\frac{dz'}{dz}\Big)^hdz^h\sum_{j=0}^{n-i-1}\frac{1}{j!}(\phi_{j+i}\circ
f)(z)\Big(
\log\Big(\frac{dz'}{dz}\Big) + \log dz\Big)^j\\
&=:& dz^h \sum_{j=0}^{n-i-1}\frac{1}{j!}\phi_{j+i}'(z)(\log dz)^j
\ee
Looking at the $(\log dz)^l$ term yields
\be
\phi_{l+i}'(z)=\Big(\frac{dz'}{dz}\Big)^h\sum_{j=0}^{n-1-i-l}\frac{1}{j!}(\phi_{j+i+l}\circ
f)(z)\Big(\log
\Big(\frac{dz'}{dz}\Big)\Big)^j
\ee
which only depends on $i+l$, rather than $i$ and $l$
independently. One thus obtains the transformation law
\be\label{logxmfn}
\phi_{i}'(z)=\Big(\frac{dz'}{dz}\Big)^h\sum_{j=0}^{n-1-i}\frac{1}{j!}(\phi_{j+i}\circ
f)(z)\Big(\log
\Big(\frac{dz'}{dz}\Big)\Big)^j
\ee
which are just the components of
\be
v'(z) = \Big( \frac{dz'}{dz}\Big)^{h+J}(v\circ f)(z)
\ee
as might be expected from (\ref{n0oneform}). Considering
$z'=z+az^{n+1}$, for $a$ infinitesimal, leads to
\be\label{loginf}
\delta \phi_i(z) := \phi'_i(z)-\phi_i(z) = a\Big( h(n+1)z^n\phi_i +
z^{n+1}\p\phi_i + (n+1)z^n\phi_{i+1}\Big)
\ee
This is the well known infinitesimal transformation law for a logarithmic primary
field \cite{rahimi}. These transformations give rise to the $n\times n$-matrix
valued vector fields
\be\label{loggydew}
l_n = \mathbb{I}(h(n+1)z^n + z^{n+1}\p ) + J(n+1)z^n\label{ln}
\ee
which act on $v$, and can be readily verified to satisfy the Witt
algebra. Just because the infinitesimal form matches up, does not
necessarily imply
that (\ref{loginf}) integrates up to (\ref{logxmfn}) by
exponentiation. This must be checked explicitly.

\section{Exponentiation}
What must be checked is that
\be
\exp (al_n)\phi_i(z) = \phi'_i(z)\label{hyp}
\ee
Since a closed form has been conjectured, this can be checked
inductively on the order of $a$. The inductive step going from
the $a^q$ to the $a^{q+1}$ is by acting on the $a^q$ term with
$\frac{a}{q+1}l_n$. The most calculationally instructive way is to build up to
this from the simplest case. Consider the case of just the co-ordinate
transformation, $z'\mapsto z$, and $l_n = z^{n+1}\frac{d}{dz}$. Then
\be\label{coord}
z' = z(1-naz^n)^{-\frac{1}{n}} = z +
\sum_{q=1}^\infty\frac{(1)(1+n)\ldots(1+n(q-1))}{q!}a^qz^{nq+1} =
z+\delta z
\ee
can be checked inductively to show that
\be
\exp (al_n)z = z(1-naz^n)^{-\frac{1}{n}}
\ee
for $n\neq 0$ (the $n=0$ case is omitted throughout, which just
corresponds to a dilation). Next, consider a function under
pull-back $g^*$, so that
\be
f'(z) = (f\circ g)(z) = f( z(1-naz^n)^{-\frac{1}{n}}) = f(z+\delta z)
\ee
Now Taylor expand in $\delta z$,
with the expression for $\delta z$ given by (\ref{coord}). The $a^q$
term is then given by
\be
f'(z)|_{a^q}=\frac{a^qz^{nq}}{q!}\sum_{k=1}^{q}\frac{(-1)^k}{k!}\sum_{p=1}^{k}\ncr{p}{k} p(p+n)\ldots(p+n(q-1))z^k\p^k\phi
\ee
In order to get the $a^{q+1}\p^{k+1}\phi$ term, one must act on the
$a^q\p^k\phi$ and $a^q\p^{k+1}\phi$ terms with
$\frac{1}{q+1}z^{n+1}\p$, and indeed the induction follows
through. For a primary field, a similar procedure can be used, where
there is now a multiplicative factor of $(\frac{dz'}{dz})^h$. Using
(\ref{coord}), one finds
\be
\Big(\frac{dz'}{dz}\Big)^h = 1+\sum_{p=1}^\infty \frac{1}{p!}h(n+1)(h(n+1)+n)\ldots(h(n+1)+n(p-1))a^pz^{np}
\ee
from which the $a^r\p^k\phi$ term can be deduced, yielding
\be
\phi'(z)|_{a^r\p^k\phi} &=&
\frac{a^rz^{nr}}{r!}(-1)^k\sum_{s=1}^k\frac{(-1)^s}{(k-s)!s!}s(s+n)\ldots
(s+n(r-1))z^k\p^k\phi +\nonumber\\
&& \frac{z^{nr}}{r!}(-1)^kh(n+1)\ldots(h(n+1)+n(p-1))\sum_{s=0}^k\frac{(-1)^s}{(k-s)!s!}z^k\p^k\phi+\nonumber\\
&&\sum_{p+q=r;\ p,q\geq
  1}\frac{a^rz^{nr}}{p!q!}h(n+1)\ldots(h(n+1)+n(p-1))\times\nonumber\\&&\qquad(-1)^k\sum_{s=1}^k\frac{(-1)^s}{(k-s)!s!}s(s+n)\ldots(s+n(q-1))z^k\p^k\phi
\ee
This can the be used in exactly the same manner as the case of the
function, with
\be
l_n = h(n+1)z^n + z^{n+1}\p
\ee
The induction is a little more involved, calculationally, than the
case of the function, but follows through in very similar
manner. Before moving to the case of the logarithmic primary field,
a way of dealing with powers of $\log$ must be found. The conjecture
\be
(-\log (1-\lambda))^k = k!\sum_{p_k =
  k}^\infty\sum_{p_{k-1}=k-1}^{p_k-1}\ldots\sum_{p_1=1}^{p_2-1}\frac{1}{p_kp_{k-1}\ldots p_1}\lambda^{p_k}
\ee
must be verified, which is easily done by induction, noting that
\be
\frac{d}{d\lambda}\Big((-\log(1-\lambda))^k\Big) =
k(-\log(1-\lambda))^{k-1}(1-\lambda)^{-1} = k(-\log(1-\lambda))^{k-1}\sum_{j=0}^\infty\lambda^j
\ee
The integration constant is fixed by noting that $(\log(1-\lambda))^k$ has
leading term $\lambda^k$. Hence, given $z'$ in (\ref{coord}),
\be
\Big(\log\Big(\frac{dz'}{dz}\Big)\Big)^k =
\Big(\frac{n+1}{n}\Big)^kk!\sum_{p_k=k}^\infty\sum_{p_{k-1}=k-1}^{p_k-1}\ldots\sum_{p_1=1}^{p_2-1}\frac{1}{p_kp_{k-1}\ldots
p_1}(naz^n)^{p_k}
\ee
For the logarithmic field, it suffices to consider the
$a^v\p^r\phi_{k+i}$ term, where $v,r,k$ are fixed, and induction
performed on them. Then,
\be
\phi'(z)|_{a^v\p^r\phi_{k+i}} =
\Big(\frac{n+1}{n}\Big)^k\Bigg[\sum_{u_{k-1}=k-1}^{v-1}\sum_{u_{k-2}=k-2}^{u_{k-1}-1}\ldots\sum_{u_1=1}^{u_2-1}\frac{n^v}{vu_{k-1}\ldots
u_1} \sum_{s=0}^r\frac{(-1)^{s+r}}{s!(r-s)!}+\nonumber\\
\Bigg(\sum_{p=1}^{v-k}\sum_{u_{k-1}=k-1}^{v-p-1}\sum_{u_{k-2}=k-2}^{u_{k-1}-1}\ldots\sum_{u_1=1}^{u_2-1}\frac{1}{(v-p)u_{k-1}\ldots
u_1}\frac{n^{v-p}}{p!}\times\nonumber\\h(n+1)\ldots
(h(n+1)+n(p-1))\sum_{s=0}^r\frac{(-1)^{s+r}}{s!(r-s)!}\Bigg) +\nonumber\\
\Bigg(\sum_{q=1}^{v-k}\sum_{u_{k-1}=k-1}^{v-q-1}\sum_{u_{k-2}=k-2}^{u_k}\ldots\sum_{u_1=1}^{u_2-1}\frac{1}{(v-q)u_k\ldots
u_1}\frac{n^{v-q}}{q!}\times\nonumber\\\sum_{s=1}^r\frac{(-1)^{r+s}}{s!(r-s)!}s(s+n)\ldots
(s+n(q-1))\Bigg) +\nonumber\\
\Bigg(\sum_{t=2}^{v-k}\sum_{u_{k-1}=k-1}^{v-t-1}\sum_{u_{k-2}=k-2}^{u_{k-1}-1}
\ldots\sum_{u_1=1}^{u_2-1}\sum_{q=1}^{t-1}\frac{1}{(v-t)u_{k-1}\ldots
u_1}\frac{n^{v-t}}{q!(t-q)!}
h(n+1)\ldots\nonumber\\
(h(n+1)+n(t-q-1))\sum_{s=1}^r\frac{(-1)^{s+r}}{s!(r-s)!}s(s+n)\ldots
(s+n(q-1))\Bigg)\Bigg] z^{nv+r}\p^r\phi_{k+i}
\ee
For the induction, the vector field $l_n$ now takes the form
(\ref{ln}). In order to find the $a^{v+1}\p^{r+1}\phi_{k+i+1}$ term,
the $a^v\p^{r+1}\phi_{k+i+1}$, $a^v\p^r\phi_{k+i+1}$ and
$a^v\p^{r+1}\phi_{k+i}$ terms must be considered. The induction is
very messy and tedious, but follows through in a very similar manner
to the previous cases. Thus, for a logarithmic field, (\ref{hyp}) is
verified, by induction. Note that nothing about whether or not $\phi$
has a Laurent expansion has been assumed, only that in a suitably small
neighbourhood, it is possible to Taylor expand $\phi$.

\section{Two Point Function}

So far, the fields $\phi_i$ have been represented by a
vector. However, as will be more useful in the following, they can be
represented as a matrix. In this instance, moving from a vector to a
matrix is analagous to moving from a vector bundle to its associated
$G$-bundle. In this sense, once given the `one-form' (\ref{gen1form}),
which generates the transformation laws,
the description of sections as matrices or vectors is equivalent. In the rank $2$ case this looks like
\be
\ncr{\phi_0}{\phi_1}\mapsto \mtr{\phi_1}{\phi_0}{0}{\phi_1}
\ee
or more generally, for a rank $N$ block, i.e. $J^N=0$, $J^{N-1}\neq 0$
\be
v \mapsto \sum_{i=0}^{N-1}\phi_i(z) J^{N-1-i} =: \mbold{\phi}(z,J)
\ee
The $l_n$ of (\ref{loggydew}) read exactly the same, except now act by
matrix multiplication, rather than by multiplication on a vector.

This notation will be useful for considering two-point
functions. Let $\mbold{\phi}(z,J)$,
$\mbold{\psi}(w,K)$ where $J^M=0$, $J^{M-1}\neq 0$ and $K^N=0$,
$K^{N-1}\neq 0$ be two logarithmic primaries. Then the two point function reads
\be
\mbold{f}(z,w,J,K) = \braket{0}{\mbold{\phi}(z,J)\otimes\mbold{\psi}(w,K)0}
\ee
where the tensor product $\otimes$ is between the vector space of
$M\times M$ matrices and $N\times N$ matrices.

One can ask what conditions the symmetry generators $l_0, l_{\pm 1}$
impose on $f$. To this end, it is useful to work in co-ordinates
$x=z-w$, $y=x+w$. The $l_{-1}$ symmetry then imposes
\be
(l_{-1}^{(1)}+l_{-1}^{(2)})\mbold{f} = 2\frac{\p}{\p y}\mbold{f} = 0
\ee
yielding $\mbold{f}=\mbold{f}(x,J,K)$. The remaining conditions then read
\be\label{loggyl0}
&(l_0^{(1)}+l_0^{(2)})\mbold{f} = \Big( \id\otimes\id (x\frac{\p}{\p x} + h_1 +
h_2 ) +J\otimes\id + \id\otimes K  \Big) \mbold{f}=0 \\ \label{loggyl1}
& (l_1^{(1)}+l_1^{(2)})\mbold{f} = \Big( y(l_0^{(1)}+l_0^{(2)}) +
x(\id\otimes\id (h_1-h_2) + J\otimes\id - \id\otimes K)   \Big) \mbold{f}=0
\ee
Since $\mbold{f}$ is a function of $J$ and $K$, it can be expanded out into a
`polynomial' in $J^m\otimes K^n$. (\ref{loggyl0}) then reads as $MN$
coupled first order ordinary differential equations in
$x$, and hence should give $MN$ independent solutions. (\ref{loggyl0}) can be rewritten, using (\ref{loggyl1}) as
\be\label{loggyl02}
& \Big( \id\otimes\id (x\frac{\p}{\p x} + 2h_1) + 2J\otimes\id \Big) \mbold{f}=0
\ee
which has solution
\be\label{gensoln1}
\mbold{f} = \mbold{C}(J,K)x^{-2(\id\otimes\id h_1 + J\otimes\id )} =
\mbold{C}(J,K)x^{-2h_1}\sum_{k=0}^{M-1} \frac{1}{k!}J^k\otimes\id (-2\log x)^k
\ee
Here, $\mbold{C}$ can be expanded as $\mbold{C} = \sum_{m,n=0}^{M-1,N-1}C_{m,n}J^m\otimes K^n$, and hence
yields $MN$ independent parameters, as required for the general solution. Using
the solution given by (\ref{gensoln1}), the
condition given by (\ref{loggyl1}) then reduces to
\be\label{loggyl12}
(h_1-h_2)C_{m,n} + C_{(m-1),n} - C_{m,(n-1)}=0
\ee
for each $m,n$ where $C_{-1,n} = C_{m,-1} = 0$. (\ref{loggyl12}) only yields
non-trivial solutions for $h_1-h_2 = 0$. Now, $M$ is not necessarily
equal to $N$, and without loss of generality, one can choose $M\leq
N$. (\ref{loggyl12}) then yields $C_{i,j} = 0$ for $i+j<N-1$. Hence,
there are only $M$ free parameters in $\mbold{C}$, which are
given by $C_{m,N-1}$.

Plugging in values for $M$ and $N$ can give rise to familiar
solutions. Consider $M=N=2$, and set $C_{1,0} = a$ and
$C_{1,1}=b$. Using a shorthand of suppressing the $\id$ and $\otimes$
symbols, one has\cite{moghi}, $\mbold{C} = (J+K)a + JKb$. Hence
\be
\mbold{C}x^{-2(h_1+J)} = x^{-2h_1}\Big( (J+K)a + JK(b-2a\log x)\Big)
\ee
Other values of $M$ and $N$ yield less familiar solutions. For an
example of different Jordan block sizes, consider $M=2$, $N=3$. One
has $\mbold{C} = (K^2 + JK)a + JK^2b$, yielding
\be
\mbold{C}x^{-2(h_1+J)} = x^{-2h_1}\Big( (K^2 + JK)a + JK^2(b-2a\log x) \Big)
\ee
To the author's knowledge, this is the first time a two-point function
has been found where the Jordan blocks differ in size.

\section{$N=1$ Logarithmic Conformal Field Theory}

The same game can be played in the $N=1$ case. 
A conformal condition in two dimensional Superconformal Field
Theory is normally specified by a
one-form being preserved up to an overall scale factor. Usually, the
preserved one-form is of the form $\om = dz - \sum_i d\g_i\g_i$. The
conformal condition then reads $f:(z,\g_i)\mapsto (z',\g_i')$ is
conformal if $f^*\om = \om\kappa$ for some function $\kappa = \kappa (z,\g_i)$, and $f$
is invertible. The invertibility condition implies that $\kappa$ has
body, and that $\frac{1}{\kappa}$ is well defined. For the $N=1$, a conformal
transformation is given by a transformation that preserves $\omega 
= dz-d\g\g =: dz +\g d\g$ up to an overall scale factor $\kappa (z,\g )$. Primary
fields can be defined \cite{nagi} as sections of the locally rank $1$ sheaf given
by $\omega^h$. Just as before, one can instead consider
$\omega^{h+J}$. Going through the same machinery, this gives the
transformation laws for $J^2=0$, and a conformal transformation $f$,
\be
&&\Phi_0'(z,\g) = (D\g ')^{2h}\Big( (\Phi_0\circ f)(z,\g) + 2\log (D\g
')(\Phi_1\circ f)(z,\g) \Big)\nonumber\\ &&\Phi_1'(z,\g) = (D\g ')^{2h}(\Phi_1\circ f)(z,\g)
\ee
which gives rise to infinitesimal transformations, with
$n\in\mathbb{Z}, r\in\mathbb{Z}+\frac{1}{2}$
\be
&&\co{L_n}{\Phi_0} = h(n+1)z^n\Phi_0 + z^{n+1}\p\Phi_0 +
\frac{n+1}{2}z^n\g\p_\g\Phi_0 + (n+1)z^n\Phi_1\nonumber\\
&&\co{G_r}{\Phi_0} = h(2r+1)z^{r-\frac{1}{2}}\g\Phi_0 +
z^{r+\frac{1}{2}}\g\p\Phi_0 - z^{r+\frac{1}{2}}\p_\g\Phi_0 +
(2r+1)z^{r-\frac{1}{2}}\g\Phi_1\nonumber\\ 
&&\co{L_n}{\Phi_1} = h(n+1)z^n\Phi_1 + z^{n+1}\p\Phi_1 +
\frac{n+1}{2}z^n\g\p_\g\Phi_1 \nonumber\\
&&\co{G_r}{\Phi_1} = h(2r+1)z^{r-\frac{1}{2}}\g\Phi_1 +
z^{r+\frac{1}{2}}\g\p\Phi_1 - z^{r+\frac{1}{2}}\p_\g\Phi_1
\ee
which are the well known commutators for a logarithmic $N=1$
Neveu-Schwarz theory
with a rank two block. These were first found in \cite{khorrami} by a different method, namely consistency with the
Jacobi identity, and requiring $L_{-1}$ to generate translations. Acting on the vacuum, and letting $(z,\g
)\rightarrow (0,0)$ gives
\be
L_0\ket{\Phi_0} = h\ket{\Phi_0} + \ket{\Phi_1},\qquad L_0\ket{\Phi_1} =
h\ket{\Phi_1},\qquad L_n\ket{\Phi_i}=0, \qquad G_r\ket{\Phi_i}=0
\ee
for $i=1,2$, and $n,r>0$.

To study the Ramond case, one might consider $\omega = dz + z\g
d\g$. This leads to
\be
&&\Phi_0'(z,\g) = \Big(\frac{z'}{z}(D\g ')^{2}\Big)^h\Big(
(\Phi_0\circ f)(z,\g) + \log \Big( \frac{z'}{z}(D\g
')^2\Big)(\Phi_1\circ f)(z,\g) \Big)\nonumber\\ &&\Phi_1'(z,\g) = \Big(\frac{z'}{z}(D\g ')^{2}\Big)^h(\Phi_1\circ f)(z,\g)
\ee
and, with $n,r\in\mathbb{Z}$,
\be
&&\co{L_n}{\Phi_0} = h(n+1)z^n\Phi_0 + z^{n+1}\p\Phi_0 +
\frac{n}{2}z^n\g\p_\g\Phi_0 + (n+1)z^n\Phi_1\nonumber\\
&&\co{G_r}{\Phi_0} = h(2r+1)z^r\g\Phi_0 +
z^{r+1}\g\p\Phi_0 - z^r\p_\g\Phi_0 +
(2r+1)z^r\g\Phi_1\nonumber\\ 
&&\co{L_n}{\Phi_1} = h(n+1)z^n\Phi_1 + z^{n+1}\p\Phi_1 +
\frac{n}{2}z^n\g\p_\g\Phi_1 \nonumber\\
&&\co{G_r}{\Phi_1} = h(2r+1)z^r\g\Phi_1 +
z^{r+1}\g\p\Phi_1 - z^r\p_\g\Phi_1
\ee
By acting on the vacuum and looking at $(z,\g)\rightarrow (0,0)$, the formulae
\be
L_0\ket{\Phi_0} = h\ket{\Phi_0} + \ket{\Phi_1},\qquad L_0\ket{\Phi_1} =
h\ket{\Phi_1},\qquad L_n\ket{\Phi_i}=0, \qquad G_r\ket{\Phi_i}=0
\ee
for $i=1,2$, and $n,r>0$ are still obtained. Now,
\be
\lim_{z,\g\rightarrow 0}\co{G_0}{\Phi_0}\ket{0} = \lim_{z,\g\rightarrow 0} \Big(h\g\Phi_0 +
z\g\p\Phi_0 - \p_\g\Phi_0 + \g\Phi_1\Big)\ket{0} = -\ket{\p_\g\Phi_0}
\ee
and hence the usual $G_0$ action on the highest weight in a Ramond
theory does not appear to be affected.

\section{$N=2$ Conformal Field Theory}

For the $N=2$ case, the preserved one-form is $\om = dz - \sum_{i=1}^2 d\g_i\g_i$.
The conformal condition reads $f:(z,\g_i)\mapsto (z',\g_i')$ is
conformal if $f^*\om = \om\kappa$ for some function $\kappa$, and $f$
is invertible. From this, it can be deduced that the superderivatives $D_i =
\frac{\p}{\p\g_i} + \g_i\frac{\p}{\p z}$ enjoy the property
$\sum_j (D_i\g_j')(D_k\g_j') = \delta_{ik}\kappa$ and hence $\frac{D_i\g_j'}{\sqrt{\kappa}}$ is
an even Grassmann complex orthogonal matrix.

In an $N=2$ Neveu-Schwarz theory, where $i=1,2$ in $\om$, the infinitesimal transformations can be represented by
the differential operators
\be
&l_m = -z^m\Big( z\frac{\p}{\p z} +
\frac{1}{2}(m+1)\g_i\frac{\p}{\p\g_i}\Big)\qquad t_m = z^m\Big(
\g_1\frac{\p}{\p\g_2} - \g_2\frac{\p}{\p \g_1} \Big)\nonumber\\
&g_r^i = z^{r-\frac{1}{2}}\Big( z\g_i \frac{\p}{\p z} -
z\frac{\p}{\p\g_i} + (r+\frac{1}{2})\g_i\g_j\frac{\p}{\p\g_j}\Big)
\ee
The $t_m$ term represents a $O(2)$ symmetry on the space of
functions. This term can be
diagonalized by the change of co-ordinates $m:(\g_1,\g_2 )\mapsto (\g^+,
\g^-)$, given by
\be
\g^+ = \frac{1}{\sqrt{2}}(\g_1 + i\g_2)\qquad \g^- =
\frac{1}{\sqrt{2}}(\g_1 - i\g_2)
\ee
This is only a change of co-ordinates, and not a conformal
transformation. The co-ordinate change amounts to studying
those transformations that preserve the one-form $\nu = dz -
d\g^+\g^- - d\g^-\g^+$. The condition for a transformation
$g:(z,\g^+,\g^-)\mapsto (z',{\g^+}',{\g^-}')$ to be conformal is
\be
g^*\nu = \nu\kappa
\ee
with $g$ invertible. Consider now a conformal (wrt $\nu$) transformation $g
:(z,\g^+, \g^- )\mapsto (z', {\g^{+}}', {\g^-}' )$ with conformal
scaling factor $\kappa$. Then $m^{-1}g^* m$
gives a conformal transformation wrt $\om$, with conformal scaling
factor $\kappa\circ m$.
\be
m^{-1}g^* m(\om(z', {\g_1}', {\g_2}')) = m^{-1}g^*\big( \nu (z', {\g^{+}}',
{\g^-}')\big) &=& m^{-1}(\nu(z, {\g^{+}}, {\g^-})\kappa\big( z, {\g^{+}},
{\g^-})\big)\nonumber\\
&=& \om(z,\g_1,\g_2 )(\kappa\circ m)(z,\g_1,\g_2 )
\ee
A similar calculation can be done starting with a conformal
transformation wrt $\om$, hence the groups of transformations for the
two conformal conditions are isomorphic. The conformal condition implies that
\be
&\kappa = \p z' + {\g^+}'\p{\g^-}' + {\g^-}'\p{\g^+}'\\
&D_+ z' = {\g^+}'D_+ {\g^-}' + {\g^-}'D_+{\g^+}'\\
&D_- z' = {\g^+}'D_- {\g^-}' + {\g^-}'D_-{\g^+}'
\ee
where $D_{\pm} = \frac{\p}{\p \g^{\pm}} + \g^{\mp}\frac{\p}{\p
  z}$. From this definition of $D_\pm$, the graded commutators $\co{D_+}{D_+}=0$,
$\co{D_-}{D_-}=0$ and $\co{D_+}{D_-}=\frac{\p}{\p z}$ can be
calculated, from which it can be seen that
\be
&(D_+{\g^+}')(D_-{\g^-}') + (D_+{\g^-}')(D_-{\g^+}') = \kappa\\
&(D_+{\g^+}')(D_+{\g^-}') = 0\\
&(D_-{\g^-}')(D_-{\g^+}') = 0
\ee
Under the conformal transformation, the superderivatives transform as
\be\label{sdev}
 \left( \begin{array}{cc} D_+ \\ D_- \end{array} \right)= \left(
 \begin{array}{cc} D_+{\g^+}' & D_+{\g^-}' \\ D_-{\g^+}' & D_-{\g^-}'
 \end{array} \right) \left( \begin{array}{cc} D_+' \\ D_-' \end{array}
 \right) = M \left( \begin{array}{cc} D_+' \\ D_-' \end{array}
 \right)
\ee
Consider now the product of matrices
\be
\left(
 \begin{array}{cc} D_+{\g^+}' & D_+{\g^-}' \\ D_-{\g^+}' & D_-{\g^-}'
 \end{array} \right) \left(
 \begin{array}{cc} D_-{\g^-}' & D_+{\g^-}' \\ D_-{\g^+}' & D_+{\g^+}'
 \end{array} \right) = \kappa \mathbb{I}_2
\ee
Taking determinants, it can be seen that $\det M=\pm\kappa$. In the
$\det M=+\kappa$ case, $\kappa = (D_+{\g^+}')(D_-{\g^-}')$. Therefore, $D_+{\g^+}'$
and $D_-{\g^-}'$ must both have body, implying
$D_-{\g^+}'=0=D_+{\g^-}'$. In the $\det M = -\kappa$ case, $\kappa =
(D_+{\g^-}')(D_-{\g^+}')$, and similarly to the previous case,
$D_+{\g^+}'=0=D_-{\g^-}'$. These two cases can be related to the $O(2)$
symmetry. Explicitly, the first case has
\be
\frac{M}{\sqrt{\kappa}} =
\mtr{\left(\frac{D_+{\g^+}'}{D_-{\g^-}'}\right)^\frac{1}{2}}{0}{0}{\left(\frac{D_+{\g^+}'}{D_-{\g^-}'}\right)^{-\frac{1}{2}}}
\ee
This gives rise to a decomposable representation of $SO(2)$. The
second case gives
\be
\frac{M}{\sqrt{\kappa}} =
\mtr{0}{\left(\frac{D_+{\g^-}'}{D_-{\g^+}'}\right)^\frac{1}{2}}{\left(\frac{D_+{\g^-}'}{D_-{\g^+}'}\right)^{-\frac{1}{2}}}{0}
\ee
which is in a region of $O(2)$ disconnected from the identity. If one
is only concerned with those transformations connected to the
identity, only the first case is of concern. Looking at only the
transformations connected to the identity is sufficient to study the
related conformal algebra. To this end, as far as the $SO(2)$ symmetry
is concerned, only the transformation rule of
$\nu^{-\frac{1}{2}}\otimes D_+ =:\delta$ is
needed. For transformations only in the connected part of the
conformal group, this locally gives a rank $1$ sheaf over a graded
Riemann sphere, and hence the restriction maps give rise to an abelian
group. Looking at the representations of this group, primary
superfields can be constructed as sections
$\mathbf{\Phi}$ of $\nu^h\otimes \delta ^q$
which, under pull-back, then yields the familiar transformation law
\be
(f^*\mathbf{\Phi})(z, \g^+,\g^-) &=& \kappa^h
\left(\textrm{$\frac{D_+{\g^+}'}{D_-{\g^-}'}$}\right) ^\frac{q}{2}(\Phi\circ
f)(z, \g^+,\g^-)\nu^h\otimes\delta ^q \nonumber\\ &=&
(D_+{\g^+}')^{h+\frac{q}{2}}(D_-{\g^-}')^{h-\frac{q}{2}}\Phi (z',
{\g^+}',{\g^-}')\nu^h\otimes\delta ^q\\ &=:&
\Phi'(z, \g^+, \g^-) \nu^h\otimes\delta ^q
\ee

\section{$N=2$ Logarithmic Conformal Field Theory}

Recall for the bosonic case, a logarithmic CFT was found by formally
replacing $h$ with $h+J$ in the exponent of $dz$. In the $N=2$ case
there are two exponents in $\nu^h\otimes \delta ^q$, $h$ and $q$, which be replaced by $h\mathbb{I}_{V_A}+A$ and
$q\mathbb{I}_{V_B}+B$ respectively, where $A$ and $B$ are nilpotent matrices on
finite dimensional vector spaces $V_A$ and $V_B$ respectively. Then a
section, $\mathbf{\Phi}$,  of
\be
\nu^{h+A}\otimes \delta^{q+B}
\ee
will be $V_A\otimes V_B$ valued. Under pull-back, one obtains
\be
f^*\mathbf{\Phi} &=& f^*(\nu^{h\mathbb{I}_{V_A}+A}\otimes \delta^{q\mathbb{I}_{V_B}+B} \Phi )
\\
&=& \left(\nu^{h\mathbb{I}_{V_A}+A}\big( (D_+\g^+)(D_-\g^-)\big)^{h\mathbb{I}_{V_A}+A}\otimes
\delta^{q\mathbb{I}_{V_B}+B}\left(
\frac{D_+\g^+}{D_-\g^-}\right)^\frac{q\mathbb{I}_{V_B}+B}{2}\right)(\Phi\circ
f)\nonumber\\
&=& \Big(\nu^{h\mathbb{I}_{V_A}+A}\otimes
\delta^{q\mathbb{I}_{V_B}+B}\Big) \left(\big( (D_+\g^+)(D_-\g^-)\big)^{h\mathbb{I}_{V_A}+A}\otimes
\left(\frac{D_+\g^+}{D_-\g^-}\right)^\frac{q\mathbb{I}_{V_B}+B}{2}\right)(\Phi\circ
f)\nonumber
\ee
This yields the transformation law
\be\label{n2log}
\Phi '(z, \g^+, \g^-) =\left(\big( (D_+\g^+)(D_-\g^-)\big)^{h\mathbb{I}_{V_A}+A}\otimes
\left(\frac{D_+\g^+}{D_-\g^-}\right)^\frac{q\mathbb{I}_{V_B}+B}{2}\right)\Phi
(z', {\g^+}', {\g^-}')
\ee
Using the infinitesimal transformations given by the superconformal
condition, (\ref{n2log}) gives the vector fields
\be
l_m \Phi &=& -z^m \Bigg( \Big( z\p + \frac{1}{2}(m+1)(\g^+\p_+ + \g^-\p_-) + 
h(m+1) - \frac{q}{2}m(m+1)\frac{\g^+\g^-}{z}\Big)\id\otimes
\id\nonumber\\
&&\qquad + (m+1)(A\otimes\id) - m(m+1)\frac{\g^+\g^-}{2z}(\id\otimes
B) \Bigg) \Phi\nonumber\\
g_{+r}\Phi &=& z^{r-\frac{1}{2}} \Bigg( \Big( z\g^-\p - z\p_+ +
(r+\frac{1}{2})\g^-\g^+\p_+ +
(2h+q)(r+\frac{1}{2})\g^-\Big)\id\otimes\id\nonumber\\
&&\qquad + \g^-(r + \frac{1}{2})(2A\otimes\id + \id\otimes B) \Bigg)\Phi\nonumber\\
g_{-r}\Phi &=& z^{r-\frac{1}{2}} \Bigg( \Big( z\g^+\p - z\p_- +
(r+\frac{1}{2})\g^+\g^-\p_- +
(2h-q)(r+\frac{1}{2})\g^+\Big)\id\otimes\id\nonumber\\
&&\qquad + \g^+(r + \frac{1}{2})(2A\otimes\id - \id\otimes B)
\Bigg)\Phi\\
j_m\Phi &=& -z^m\Bigg( \Big(\g^+\p_+ - \g^-\p_- - 2mh\frac{\g^+\g^-}{z}
+ q\Big)\id\otimes\id - 2m\frac{\g^+\g^-}{z}(A\otimes\id) + (\id\otimes B)\Bigg)\Phi\nonumber
\ee
These vector fields then obey the graded commutation relations of the
centreless $N=2$ algebra
\be
&\co{l_m}{l_n} = (m-n)l_{m+n}\qquad \co{l_m}{g_{\pm
    r}}=(\frac{m}{2}-r)g_{\pm(m+r)}\qquad
\co{l_m}{t_n}=-nj_{m+n}\nonumber\\&\nonumber\\&
\co{g_{+r}}{g_{-s}}=2l_{r+s}+(r-s)j_{r+s}\qquad \co{g_{\pm r}}{g_{\pm
    s}}=0\nonumber\\&\nonumber\\&
\co{j_m}{g_{\pm r}} = \pm g_{\pm (m+r)}\qquad \co{j_m}{j_n}=0
\ee

\section{Two Point Function}

Consider first the case $A=0=B$. Then the symmetry generators can be
used to calculate
\be
f(Z_1, Z_2) = f(z, \g_1^+, \g_1^-, w, \g_2^+, \g_2^- ) =
\langle\Phi_1(Z_1)\Phi_2(Z_2) \rangle.
\ee
Consider the change of variables
\be
&Z_{12} = (z-w)-(\g_1^+\g_2^- + \g_1^-\g_2^+),\qquad W_{12} =
(z+w)-(\g_1^+\g_2^- + \g_1^-\g_2^+)\nonumber\\
&\g_{12}^+ = \g_1^+ - \g_2^+,\qquad \g_{12}^- = \g_1^- - \g_2^-, \qquad
\xi_{12}^+ = \g_1^+ + \g_2^+,\qquad \xi_{12}^- = \g_1^- + \g_2^-
\ee
Then
\be
&\frac{\p}{\p Z_{12}} = \frac{1}{2}\Big(\frac{\p}{\p z}-\frac{\p}{\p
w}\Big), \qquad \frac{\p}{\p W_{12}} = \frac{1}{2}\Big(\frac{\p}{\p z}+\frac{\p}{\p
w}\Big)\nonumber\\
&\frac{\p}{\p\g_{12}^+} = \frac{1}{2}\Big(\frac{\p}{\p\g_1^+} -
\frac{\p}{\p\g_2^+} +(\g_2^- + \g_1^-)\frac{\p}{\p z}\Big), \qquad \frac{\p}{\p\g_{12}^-} = \frac{1}{2}\Big(\frac{\p}{\p\g_1^-} -
\frac{\p}{\p\g_2^-} +(\g_2^+ + \g_1^+)\frac{\p}{\p z}\Big)\nonumber\\
&\frac{\p}{\p\xi_{12}^+} = \frac{1}{2}\Big(\frac{\p}{\p\g_1^+} +
\frac{\p}{\p\g_2^+} +(\g_2^- - \g_1^-)\frac{\p}{\p z}\Big), \qquad \frac{\p}{\p\xi_{12}^-} = \frac{1}{2}\Big(\frac{\p}{\p\g_1^-} +
\frac{\p}{\p\g_2^-} +(\g_2^+ - \g_1^+)\frac{\p}{\p z}\Big)
\ee
Now consider the action of the lie algebra on $f$.
\be\label{n2l-1cndn}
&(l_{-1}^{(1)} + l_{-1}^{(2)})f = 0 = \Big(\frac{\p}{\p z}+\frac{\p}{\p
w}\Big) f = 2\frac{\p}{\p W_{12}}f\nonumber\\
&(g_{+,-\frac{1}{2}}^{(1)} + g_{+,-\frac{1}{2}}^{(2)})f = 0 =
\Big(\frac{\p}{\p\g_1^+} - \g_1^-\frac{\p}{\p z}+
\frac{\p}{\p\g_2^+} -\g_2^-\frac{\p}{\p w}\Big) f=
2\frac{\p}{\p\xi_{12}^+} f\nonumber\\
&(g_{-,-\frac{1}{2}}^{(1)} + g_{-,-\frac{1}{2}}^{(2)})f = 0 =
\Big(\frac{\p}{\p\g_1^-} - \g_1^+\frac{\p}{\p z}+
\frac{\p}{\p\g_2^-} -\g_2^+\frac{\p}{\p w}\Big) f=
2\frac{\p}{\p\xi_{12}^-} f
\ee
Hence $f=f(\g_{12}^+, \g_{12}^-, Z_{12})$. Similarly
\be\label{n2l0cndn}
&&(l_0^{(1)} + l_0^{(2)})f = 0 =\\ &&\Big( h_1 +h_2 + Z_{12}\frac{\p}{\p Z_{12}} +
W_{12}\frac{\p}{\p W_{12}} + \frac{1}{2}\Big(\g_{12}^+\frac{\p}{\p\g_{12}^+} +
\xi_{12}^+\frac{\p}{\p\xi_{12}^+} +\g_{12}^-\frac{\p}{\p\g_{12}^-}
+\xi_{12}^-\frac{\p}{\p\xi_{12}^-} \Big)\Big) f\nonumber
\ee
yielding
\be
f = a_0Z_{12}^{-h_1-h_2} + a_+ \g_{12}^+
Z_{12}^{-h_1-h_2-\frac{1}{2}} + a_- \g_{12}^- Z_{12}^{-h_1-h_2-\frac{1}{2}} + a_{+-}\g_{12}^+\g_{12}^-Z_{12}^{-h_1-h_2-1}
\ee
where $a_0$, $a_{+-}$ are graded even constants, and $a_\pm$ are
graded odd constants. Applying the condition
\be\label{n2j0cndn}
&(j_0^{(1)}+j_0^{(2)})f = 0 = \Big( q_1+q_2+
\g_{12}^+\frac{\p}{\p\g_{12}^+} + \xi_{12}^+\frac{\p}{\p\xi_{12}^+} -
\g_{12}^-\frac{\p}{\p\g_{12}^-} - \xi_{12}^-\frac{\p}{\p\xi_{12}^-}
\Big) f
\ee
yields the solutions either $(q_1+q_2)=0=a_\pm$, or
$(q_1+q_2+1)=0=a_0=a_{+-}=a_{-}$, or
$(q_1+q_2-1)=0=a_0=a_{+-}=a_{+}$. Since the commutator of $l_1$ with
$g_{\pm ,-\frac{1}{2}}$ gives $g_{\pm ,\frac{1}{2}}$, only the $l_1$
condition need be applied.
\be\label{n2l1cndn}
&&(l_1^{(1)}+l_1^{(2)})f = 0 =\nonumber\\ && \frac{1}{4}\Big\lbrack (W_{12}^2+Z_{12}^2)\frac{\p}{\p W_{12}} +
4W_{12}Z_{12}\frac{\p}{\p Z_{12}} + 2Z_{12}(\g_{12}^+\xi_{12}^- +
\g_{12}^-\xi_{12}^+)\Big(\frac{\p}{\p Z_{12}}+\frac{\p}{\p
W_{12}}\Big)\nonumber\\&&
-\xi_{12}^+(\xi_{12}^- + \g_{12}^-)\g_{12}^+\Big(\frac{\p}{\p
\xi_{12}^+}+\frac{\p}{\p \g_{12}^+}\Big) -\xi_{12}^-(\xi_{12}^+ + \g_{12}^+)\g_{12}^-\Big(\frac{\p}{\p
\xi_{12}^-}+\frac{\p}{\p \g_{12}^-}\Big)\nonumber\\&&
2W_{12}(\xi_{12}^+\frac{\p}{\p\xi_{12}^+} +
\g_{12}^+\frac{\p}{\p\g_{12}^+} + \xi_{12}^-\frac{\p}{\p\xi_{12}^-} +
\g_{12}^-\frac{\p}{\p\g_{12}^-}) + \nonumber\\ &&2Z_{12}(\xi_{12}^+\frac{\p}{\p\g_{12}^+} +
\g_{12}^+\frac{\p}{\p\xi_{12}^+} + \xi_{12}^-\frac{\p}{\p\g_{12}^-} +
\g_{12}^-\frac{\p}{\p\xi_{12}^-}) + 4W_{12}(h_1+h_2) +
4Z_{12}(h_1-h_2) +\nonumber\\ && 4h_1(\g_{12}^+\xi_{12}^- +
\g_{12}^-\xi_{12}^+) - (q_1+q_2)(\xi_{12}^+\xi_{12}^- +
\g_{12}^+\g_{12}^-) - (q_1-q_2)(\g_{12}^+\xi_{12}^- +
\xi_{12}^+\g_{12}^-)\Big\rbrack f = \nonumber\\
 && \frac{1}{4}\Big\lbrack 4W_{12}(l_0^{(1)}+l_0^{(2)}) + 2Z_{12}(\g_{12}^+\xi_{12}^- +
\g_{12}^-\xi_{12}^+)\Big(\frac{\p}{\p Z_{12}}\Big)
-\xi_{12}^+(\xi_{12}^- + \g_{12}^-)\g_{12}^+\Big(\frac{\p}{\p
\g_{12}^+}\Big) -\nonumber\\ &&\xi_{12}^-(\xi_{12}^+ + \g_{12}^+)\g_{12}^-\Big(\frac{\p}{\p \g_{12}^-}\Big)
 +  2Z_{12}(\xi_{12}^+\frac{\p}{\p\g_{12}^+} +
 \xi_{12}^-\frac{\p}{\p\g_{12}^-}) +
4Z_{12}(h_1-h_2) +\nonumber\\ && 4h_1(\g_{12}^+\xi_{12}^- +
\g_{12}^-\xi_{12}^+) - (q_1+q_2)(\xi_{12}^+\xi_{12}^- +
\g_{12}^+\g_{12}^-) - (q_1-q_2)(\g_{12}^+\xi_{12}^- +
\xi_{12}^+\g_{12}^-)\Big\rbrack f \nonumber\\ = 
 && \frac{1}{4}\Big\lbrack 2(2W_{12}+\g_{12}^+\xi_{12}^- +
\g_{12}^-\xi_{12}^+)(l_0^{(1)}+l_0^{(2)}) - \nonumber\\ &&(\xi_{12}^+\xi_{12}^- +
\g_{12}^+\g_{12}^- + \g_{12}^+\xi_{12}^- + \xi_{12}^+\g_{12}^-)(j_0^{(1)}+j_0^{(2)})
+2(h_1-h_2)(2Z_{12}+\g_{12}^+\xi_{12}^- + \nonumber\\ &&\g_{12}^-\xi_{12}^+) +
2Z_{12}(\xi_{12}^+\frac{\p}{\p\g_{12}^+} +
\xi_{12}^-\frac{\p}{\p\g_{12}^-}) - (\g_{12}^+\xi_{12}^- +
\g_{12}^-\xi_{12}^+)(\g_{12}^+\frac{\p}{\p\g_{12}^+}
+\g_{12}^-\frac{\p}{\p\g_{12}^-}) + \nonumber\\ &&2q_2(\g_{12}^+\xi_{12}^- +
\xi_{12}^+\g_{12}^-) \Big\rbrack f
\ee
which then yields only one possibly non-trivial solution, namely
$h_1-h_2=q_1+q_2=0=a_\pm$ and $a_{+-}=-q_2a_0$.

Consider now non-zero $A,B$, which amounts to replacing $h_1$ by
$h_1+J$, $h_2$ by $h_2+K$, $q_1$ by $q_1+P$ and $q_2$ by $q_2+Q$. The extra
parameters have the properties that $J^M, K^N, P^R, Q^S=0$ and
$J^{M-1}, K^{N-1}, P^{R-1}, Q^{S-1}\neq 0$. So as not to overly
clutter the notation, tensor product signs will be omitted. (\ref{n2l-1cndn}) remains
unchanged, and hence $\mbold{f}=\mbold{f}(\g_{12}^+, \g_{12}^-,
Z_{12},J,K,P,Q)$. (\ref{n2l0cndn}) is modified to
\be
\Big( h_1+J+h_2+K+Z_{12}\frac{\p}{\p Z_{12}} +
\frac{1}{2}\Big(\g_{12}^+\frac{\p}{\p\g_{12}^+} +
\g_{12}^-\frac{\p}{\p\g_{12}^-}\Big) \Big) \mbold{f} = 0
\ee
which, similarly to the bosonic case, has solution
\be
\mbold{f} = \mbold{a}_0Z_{12}^{-\Delta} +
\mbold{a}_+\g_{12}^+Z_{12}^{-\Delta-\frac{1}{2}}+
\mbold{a}_-\g_{12}^-Z_{12}^{-\Delta-\frac{1}{2}}+
\mbold{a}_{+-}\g_{12}^+\g_{12}^-Z_{12}^{-\Delta-1}
\ee
where $\Delta = h_1+J+h_2+K$, and the prefactors have dependence
$\mbold{a} = \mbold{a}(J,K,P,Q)$. (\ref{n2j0cndn}) becomes
\be
\Big( q_1+ P + q_2+ Q+
\g_{12}^+\frac{\p}{\p\g_{12}^+} -
\g_{12}^-\frac{\p}{\p\g_{12}^-}
\Big) \mbold{f}=0
\ee
This yields the conditions on the prefactors
\be
&(q_1+q_2+P+Q)\mbold{a}_{0}=0,\qquad
(q_1+q_2+P+Q+1)\mbold{a}_{+}=0,\nonumber\\ &
(q_1+q_2+P+Q-1)\mbold{a}_{-}=0,\qquad (q_1+q_2+P+Q)\mbold{a}_{+-}=0
\ee
Now, since $(P+Q)$ is nilpotent, if $q_1+q_2\neq 0,\pm 1$, the above
relations can be inverted to show that all the $\mbold{a}=0$. The
possibly non-trivial solutions are given by
\be
q_1+q_2=0 &\Rightarrow &\mbold{a}_{\pm}=0,\quad
(P+Q)\mbold{a}_0=0,\quad (P+Q)\mbold{a}_{+-}=0\\
q_1+q_2=1 &\Rightarrow &\mbold{a}_0=\mbold{a}_+=\mbold{a}_{+-}=0,\quad
(P+Q)\mbold{a}_-=0\\
q_1+q_2=-1 &\Rightarrow &\mbold{a}_0=\mbold{a}_-=\mbold{a}_{+-}=0,\quad
(P+Q)\mbold{a}_+=0
\ee 
(\ref{n2l1cndn}) now reads
\be
\Big( 2(h_1-h_2+J-K)(2Z_{12}+\g_{12}^+\xi_{12}^- + \g_{12}^-\xi_{12}^+) +
2(q_2+Q)(\g_{12}^+\xi_{12}^- + \xi_{12}^+\g_{12}^-) +\nonumber\\
2Z_{12}\Big(\xi_{12}^+\frac{\p}{\p\g_{12}^+} +
\xi_{12}^-\frac{\p}{\p\g_{12}^-}\Big) -
(\g_{12}^+\xi_{12}^-+\g_{12}^-\xi_{12}^+)\Big(
\g_{12}^+\frac{\p}{\p\g_{12}^+} +
\g_{12}^-\frac{\p}{\p\g_{12}^-}\Big)\Big) \mbold{f}=0
\ee
For $q_1+q_2=\pm 1$, this yields only the trivial solution, and for
$q_1+q_2=0$ yields $(q_2+Q)\mbold{a}_0 + \mbold{a}_{+-} = 0 = h_1-h_2$ and
$(J-K)\mbold{a}_0=0$. Hence
\be
\mbold{f} = \mbold{a}_0\Big( Z_{12}^{-\Delta} -
(q_2+Q)\g_{12}^+\g_{12}^-Z_{12}^{-\Delta -1}\Big)
\ee
subject to $(J-K)\mbold{a}_0 = 0 = (P+Q)\mbold{a}_0$. $\mbold{a}_0$
thus has $\min(M,N)\times\min(R,S)$ free parameters. In particular,
allowing $j_0$ to be non-diagonalizable does not seem to introduce any
extra logarithms into the two point function. This is perhaps not too
surprising, since in the two point function, $Q$ only appears in conjunction
with nilpotent variables.

\section{Conclusions}
Bosonic Logarithmic CFT was studied from a more geometric point of
view, yielding familiar results of the definition of a logarithmic
primary field. In particular, an example was found of a two-point
function, where the two Jordan blocks were of different size, that was
not set to zero by the global conformal invariance. The generators of
the infinitesimal transformation were shown to
integrate up to the geometric field defined. Using the machinery
developed, the two-point function was obtained. The construction
was applied to the $N=1$ case, where again familiar results were
found. The construction was then applied to the $N=2$ case, and the
two point function calculated. The only logarithmic divergences
occurred in a manner familiar to the bosonic and $N=1$ cases, even
though the Cartan subalgebra is enlargened, and contains more
non-diagonalizable elements than the bosonic and $N=1$ cases.

Despite the fact that considering $\log dz$ has been useful in
constructing LCFTs, precisely what it is is still not obvious to the
author. It does not seem like something that could exist on a Riemann
Sphere, perhaps some kind of covering of the sphere is needed. What is still an
intriguing question is - when demanding this type of non-unitary
behaviour, what are the implications for the geometry of the
underlying space on which the CFT is built?

Precisely how to generalize the machinery found in this note to $N=3$
superconformal theories is not entirely
obvious. The $R$-symmetry at the lie algebra level is given by $su(2)$, and is
non-abelian. Presumably, representations of $su(2)$ where $J_3$ is
non-diagonalizable would be required.

\subsection*{Acknowledgements}

The author would like to thank Michael Flohr and J{\o}rgen Rasmussen
for various comments and discussion over e-mail.

\end{document}